\documentstyle[12pt]{article}

\newcommand{\LL}{{I\!\! L}}
\newcommand{\eq}[1]{(\ref{#1})}
\newcommand{\diff}{\partial}
\newcommand{\beq}{\begin{equation}}
\newcommand{\eeq}{\end{equation}}
\newcommand{\beqn}{\begin{eqnarray}}
\newcommand{\eeqn}{\end{eqnarray}}
\newcommand{\cD}{{\cal D}}

\def\dd{{\rm d}}
\def\NP{ Nucl.~Phys.}
\def\PR{ Phys.~Rev.}
\def\PL{ Phys.~Lett.}
\def\PRL{ Phys.~Rev.~Lett.}

\begin{document}
\centerline{\bf~~~~~~~~~~~~~~~~~~~~~~~~~~~~~~~~~~~~~~~~~~~~~~~
~~~~~~~~~~~~~~~~~~~~~~~~~~ITEP--TH--16/96}

\vspace{20mm}

\centerline{\bf \Large Fermionic String from Abelian Higgs Model with}
\centerline{\bf \Large monopoles and $\Theta$-term}

\vspace{10mm}

\centerline{E.T.~Akhmedov\footnote{e--mail: akhmedov@vxitep.itep.ru}}

\centerline{Institute of Theoretical and Experimental Physics}

\centerline{Moscow, 117259, B. Cheremushkinskaya, 25.}

\vspace{10mm}

\begin{abstract}

The four dimensional Abelian Higgs model with monopoles and $\Theta$-term is
considered in the limit of the large mass of the higgs boson. We show that
for $\Theta=2 \pi$ the theory is equivalent, at large distances, to
summation over all possible world-sheets of fermionic strings with Dirichlet
type boundary conditions on string coordinates.

\end{abstract}

\vspace{10mm}

\centerline{\bf Dedicated to the memory of Alexandr Hohlov.}

\newpage

   There are processes in quantum field theory in which a particle
description is not convenient. The examples are QCD at low energies
\cite{Pol87} and some astrophysical phenomena in early universe
\cite{HiKi94}.  Due to that it is interesting to study string theories
which follow from field theories.

   For example, the quantum theory of Abrikosov--Nielsen--Olesen (ANO)
strings can be obtained from the Abelian Higgs model (AHM)
\cite{PoWiZu93,Orl94,SaYa95,AkChPoZu96}. It occurs that for thin ANO strings
the theory is local. The effective action for thin ANO strings contains a
rigidity term \cite{Pol86}  with the negative sign.\footnote{The sign of the
rigidity term depends on the type (first or second) of the superconductor
\cite{Orl94}.} The string theory with Nambu--Goto and rigidity term may have
problems with unitarity, presence of tachyon in the spectrum and crumpling of
the string world--sheet (see e.g. review \cite{Dav90}).  But such problems
are absent for the Neveu--Schwarz--Ramond (NSR) string in ten
dimensions.  This string is equivalent (in the stable point of the
$\beta$--function for the rigidity term) \cite{Wie89} to the string
theory with the action containing the rigidity term plus topological
Wess--Zumino--Novikov--Witten (WZNW) term. In the present paper we give an
example how a similar theory can be obtained from the four--dimensional AHM
with monopoles and $\Theta$--term in the limit of the big mass of the Higgs
boson.

   We start with the following partition function in the Euclidian space
time\footnote{Below we assume a lattice regularization \cite{PoWiZu93} and,
in formula \eq{bb}, we take the naive continuum limit.}:

\beqn
Z = \int [\cD\tilde z_{\mu}]\cD A_{\mu}\cD \Phi \exp \biggl\{ -
\int \dd^4 x \biggl[ \frac{1}{4} \biggl(F_{\mu\nu} + \bar
F_{\mu\nu}(\tilde z)\biggr)^2 \nonumber \\ + \frac{1}{2} |D_{\mu} \Phi|^2 +
\lambda (|\Phi|^2 - \zeta^2)^2 + \nonumber\\ + i \frac{\Theta e^2}{32\pi^2}
\epsilon_{\mu\nu\alpha\beta}\biggl(F_{\mu\nu} + \bar F_{\mu\nu}(\tilde
z)\biggr) \biggl(F_{\alpha\beta} + \bar F_{\alpha\beta}(\tilde
z)\biggr)\biggr] \biggr\}, \label{Initial}
\eeqn
where
\beqn
D_{\mu} = \diff_{\mu} - i e A_{\mu} - i e \bar A_{\mu}, \nonumber \\
\epsilon_{\mu\nu\alpha\beta} \diff_{\nu}\bar F_{\alpha\beta}(\tilde z) =
\frac{4\pi}{e}\int_{C} \dd\tilde z_{\mu} \delta^{(4)}(x - \tilde z), \quad
\diff_{[\mu} \bar A_{\nu]} = \bar F_{\mu\nu},
\eeqn
and $j_{\mu} = \frac{1}{4\pi} \epsilon_{\mu\nu\alpha\beta} \diff_{\nu}\bar
F_{\alpha\beta}(\tilde z)$ is the conserving monopole's current:
$\diff_{\mu}j_{\mu} = 0$, $\tilde z_{\mu}$ is a position of the monopole;
$\int [\cD\tilde z_{\mu}]$ is the functional integral over all closed paths,
the measure is well known, see e.g. \cite{Pol87}; $C$ are the trajectories of
the monopoles defined by $\tilde z_{\mu}$.  $\Phi = |\Phi| e^{i \theta}$ is
the Higgs field with the standard integration measure: $\cD\Phi = \cD Re\Phi
\cD Im\Phi = [|\Phi|\cD |\Phi|] \cD\theta$.

   The theory \eq{Initial}  can be considered as the low energy
limit of the $SU(2)$ Georgy--Glashow model with the $\Theta$--term and with
the additional breaking of the gauge $U(1)$ symmetry. This model is known to
have ANO strings and 't~Hooft--Polyakov monopoles as the solutions of the
classical equations of motion \cite{HiKi94}. At the low energy, monopoles can
be considered as Wu--Yang type ambiguities in the gauge potential $A_{\mu}$
\cite{Pol87}. In \eq{Initial} we explicitly write these ambiguities as $\bar
F_{\mu\nu}(\tilde z)$.

  Since in the center of the ANO strings $Im\Phi = Re\Phi = 0$
the phase $\theta$ is singular on the two dimensional surfaces, which are
world--sheets of ANO strings. The character of the singularity is:

\beqn
\diff_{[\mu,} \diff_{\nu]} \theta^s (x, \tilde x) & = & 2 \pi \epsilon_{\mu
\nu \alpha \beta} \Sigma_{\alpha \beta}(x, \tilde x), \nonumber \\
\Sigma_{\alpha \beta}(x, \tilde x) & = &
\int_{\Sigma + \Sigma_C} \dd^2 \sigma \epsilon^{ab} \diff_a\tilde
x_{\alpha}\diff_b\tilde x_{\beta}\delta^{(4)} [x - \tilde x(\sigma)],
\label{ssss}
\eeqn
where $\Sigma$ and $\Sigma_C$ are collections of all closed surfaces and
surfaces opened on monopole's world--lines $C$.

   Using the Bianci identity ($\epsilon_{\mu\nu\alpha\beta}\diff_{\nu}
F_{\alpha\beta} = 0$) and conservation of the  \\ monopole current
($\diff_{\mu}j_{\mu} = 0$) we can rewrite the $\Theta$--term as  $\frac{\Theta
e^2}{2 \pi} j_{\mu} A_{\mu}$ \cite{Wit79}, which is the interaction of the
electric charge of the dyon with the gauge field.

   In eq.  \eq{Initial} $\cD \theta$ contains the integration over functions
which are singular on two--dimensional manifolds \eq{ssss}, and we subdivide
$\theta$ into the regular $\theta^r$ and the singular $\theta^s$ parts:
$\theta = \theta^r + \theta^s$; $\theta^s$ is defined by eq.  \eq{ssss}. To
simplify the calculations we consider the London limit
($\lambda >> 1$),  in this case the radial part $|\Phi|$ of the
Higgs field $\Phi$ is fixed $|\Phi| = \zeta$ and $\cD\theta =
\cD\theta^r\cD\theta^s$. After the change of variables from $\theta^s$ to
$\tilde x_{\mu}$ and integration over $A_{\mu}$ and $\theta^r$ in
\eq{Initial}, we get \cite{AkChPoZu96}:

\newpage
\beqn
Z = const \cdot \int [\cD \tilde z_{\mu}] [\cD \tilde x] \cdot J(\tilde x)
\cdot \nonumber \\ \exp \Biggl\{ -  \int \dd^4 x \int \dd^4
y \biggl [ \pi^2 \zeta^2 \Sigma_{\mu\nu}(x) \cD_m^{(4)}(x - y) \Sigma_{\mu
\nu}(y) + \nonumber \\ + \biggl(\biggl(\frac{\Theta e}{2\pi}\biggr)^2 +
\frac{1}{4 e^2}\biggr) \cdot j_\mu(x) \cD^{(4)}_m(x - y)j_\mu(y) + \nonumber
\\ + \frac{\Theta e}{2\pi} \cdot j_\mu(x) \cD^{(4)}_m(x - y)
\diff_{\nu}\epsilon_{\mu \nu \alpha \beta} \Sigma_{\alpha \beta}(y) \biggr ]
+ \nonumber \\ + i\Theta \cdot \LL(\Sigma, C) + i\Theta \cdot \LL(\Sigma_C,
 C)\Biggr\}\,, \label{bb}
\eeqn
where $\tilde x_{\mu}$ is the position of the string, $[\cD\tilde x_{\mu}]$
assumes both integration over all possible positions and summation over all
topologies of the string's world--sheets $\Sigma$ and $\Sigma_C$;
$\cD_m^{(4)}(x - y)$ is the Green's function: $(\Delta + m^2) \cD_m^{(4)}(x -
y) = \delta^{(4)}(x - y)$, $m^2 = e^2 \zeta^2$ is the mass of the gauge
boson; $J(\tilde x)$ is the Jacobian of the transformation from the field
$\theta^s$ to the string position $\tilde x_{\mu}$. $J(\tilde x)$ was
estimated in \cite{AkChPoZu96} for string with the topology of a sphere or of
a disk. This Jacobian contains a term which cancels the conformal anomaly
coming from Nambu--Goto action in four dimensions. The mechanism of
cancelation is the same as suggested in \cite{PoSt91}.

  Boundary condition for the open strings in theory \eq{bb} is:
$\diff_{\mu} \Sigma_{\mu\nu} = j_{\nu}$, which leads to the Dirichlet
condition $\tilde x_{\mu}(s)|_{C} = \tilde z_{\mu}(s)$.

  First three terms in the exponent in eq. \eq{bb} describe the interaction
and the self interaction of strings and dyons through exchange of the massive
gauge bosons. The last two terms: $\LL(\Sigma, C)$ and $\LL(\Sigma_C, C)$
describe the topological interaction of strings and dyons.

\beqn
\LL(\Sigma, C) = \frac{1}{4\pi^2} \int_{C} \dd \tilde
z_{\alpha} \int_{\Sigma} \dd^2\sigma \cdot \epsilon^{ab}\cdot
\diff_a\tilde x_{\mu}\diff_b\tilde x_{\nu}\cdot \epsilon_{\mu\nu\alpha\beta}
\diff_{\beta} \frac{1}{|\tilde x - \tilde z|^2} \label{Link4D}
\eeqn
is the four--dimensional Gauss linking number of the world--sheet of the
closed string $\Sigma$ and of the dyon path $C$. This term is a
four--dimensional analogue of the Aharonov--Bohm interaction of the strings
and dyons discussed in \cite{AkChPoZu96,AlWi89}. The string behaves like a
solenoid which scatter the dyon.  The other topological interaction

\beqn
 \LL(\Sigma_C, C) = \frac{1}{4\pi^2} \int_{C} \dd \tilde z_{\alpha}
 \int_{\Sigma_C} \dd^2\sigma \cdot\epsilon^{ab}\cdot\diff_a\tilde
 x_{\mu}\diff_b\tilde x_{\nu}\cdot \epsilon_{\mu\nu\alpha\beta} \diff_{\beta}
 \frac{1}{|\tilde x - \tilde z|^2} \label{LnD}
\eeqn
is the generalization to the four dimensions of a similar
interaction  of open paths of the particles in three dimensions
\cite{Pol88}.  Formally it is equal to zero because there is no linking
between open surface and closed path. But since the dyon trajectory $C$
coincides with the boundary of the world--sheet of the ANO string $\Sigma_C$,
the integral in \eq{LnD} should be regularized. At each point of the
curve $C$ we define  a tangent vector $e^1_{\mu}(s^1) = \diff_{s^1}
\tilde y_{\mu}(s^1)$; the vector $e^2_{\mu}(s^1)$ which is orthogonal to
$e^1_{\mu}(s^1)$ and tangent to the surface $\Sigma_C$,  and two vectors
$n^a_{\mu}(s^1)\,\, a = 1, 2$ which are orthogonal to the $e^a_{\mu}(s^1)$
\cite{FuGaTr89}. Consider the path  $C_{\epsilon}$ which
is the shift  of the path $C$ along one of the normals (for
example\footnote{The choice of normals is unimportant even in the case of
open strings \cite{FuGaTr89}.} $n^1_{\mu}(s^1)$) from the border of the
surface $\Sigma_c$ by a distance $\epsilon$. Now we define
$\LL(\Sigma_C, C) = \lim_{\epsilon\rightarrow 0} \LL(\Sigma_C,
C_{\epsilon})$, and it is easy to find:

\beq
\LL(\Sigma_C, C) = - \frac{1}{4\pi}\int^L_0\dd s^1 \cdot
\epsilon_{\mu\nu\alpha\beta} \cdot \diff_{s^1} n^1_{\alpha}(s^1)\cdot
e^1_{\mu}(s^1)\cdot e^2_{\nu}(s^1) \cdot n^1_{\beta}(s^1) \,, \label{nnn}
\eeq
where $L$ is the length and $s^1$ is a parametrization of
the boundary $C$. If in \eq{nnn} we consider the closed surface $\Sigma$
and the path $C$ is lying on this surface then the expression in the $RHS$
can be easily represented in WZNW form \cite{FuGaTr89}. To consider the
open surface $\Sigma_C$ it is convenient to introduce (as it was done in
\cite{AlSh88}) the following three form:

\beq
\Omega_{ijk} = \epsilon_{\mu\nu\alpha\beta} e_{\mu}\diff_i
e_{\nu}\diff_j e_{\alpha}\diff_k e_{\beta}, \label{form}
\eeq
defined on some three dimensional compact manifold $B$ with boundary
containing the surface $\Sigma_C$. In the last expression $e_{\mu}$ is an
extension, to the manifold $B$, of the vector $e_{\mu}(s^1,s^2) =
\cos{(s^2)} \cdot e^1_{\mu}(s^1) + \sin{(s^2)} \cdot e^2_{\mu}(s^1)$, tangent
to $\Sigma_C$ (later we assume the similar extension of the vectors
$n^a_{\mu}$); $\diff_i = \frac{\diff}{\diff s^i}$, $i,j,k = 1,2,3$ and $s^3$
is the additional to $s^a$ coordinate on $B$.  Since $(e_{\mu})^2 =
1$ by construction, the three form \eq{form} is closed, $\diff_{[l}
\Omega_{ijk]} = 0$, and one can represent $\Omega_{ijk}$ locally as
$\diff_{[k} \Lambda_{ij]} = - \Omega_{ijk}$ where $\Lambda_{ij}$ is a
squesymmetric two form.  So that $\LL(\Sigma_C, C)$ is:

\beq
\LL(\Sigma_C, C) = \frac{1}{8\pi^2}\int_{\Sigma_C}\dd^2 s \epsilon^{ab}
\Lambda_{ab}, \quad a,b = 1,2  \label{WZWL}
\eeq
$\Lambda_{ab}$ is defined up to the transformation
$\Lambda_{ab}\rightarrow\Lambda_{ab} + \diff_{[a} E_{b]}$. But
since $\Sigma_C$ is an open surface the expression \eq{WZWL} is not
invariant under this transformation. This ambiguity affects only a boundary
terms of the string:  dyon's theory.  But our further
discussion is independent on such terms.

   Consider the part of the string theory \eq{bb} corresponding to a surface
$\Sigma_C$ with the topology of the disc.  If $e^2 > \lambda$ this
string theory is local and contains the rigidity term with the positive sine
\cite{Orl94}, which is important for the consistency of the quantum string
theory \cite{Pol86}:

\beqn
S(\Sigma_C) =  \eta \int_{\Sigma_C} \sqrt{g} \dd^2 s + k
\int_{\Sigma_C}\sqrt{g}\biggl(\Delta(g) \tilde x_{\mu}\biggr)^2 \dd^2 s
- \nonumber \\ - i\Theta \LL(\Sigma_C, C) - \ln{J(\tilde x)} +
O(\frac{1}{m^2}), \label{sia}
\eeqn
here string tension $\eta$ and rigidity $k > 0$ are some coefficients
\cite{Orl94,AkChPoZu96}.

   The WZNW term $\LL(\Sigma_C,C)$ is defined by \eq{WZWL} for a particular
choice of the reference system $e^a_{\mu}$ and $n^a_{\mu}$ (parametrization
of the world--sheet $\Sigma_C$).  To get this term for an arbitrary reference
system we can rotate it to any other position by some $SO(4)$ matrix.  The
last matrix is defined up to $SO(2)$ local rotations on the string
world--sheet and up to $SO(2)$ local rotations in the space orthogonal to the
world--sheet. The last $SO(2)$ invariance is defined by the following matrix
$h$ \cite{Wie89}: consider the gauge field $A^n = n_{\mu}^a \dd n_{\mu}^b
\cdot M^{ab}$, where $M^{ab}$ is the generator of the $SO(2)$ rotations. The
definition of $h$ is the following: $\dd_{+}(h^{-1}\dd_{-}h) = \dd A^n + A^n
\wedge A^n$.  By this way the theory \eq{sia} can be represented as a gauged
$\frac{SO(4)}{SO(2) \cdot SO(2)}$ WZNW model, which is related to the
geometrical quantization on the group orbits.  In this model the
$\beta$--function for the rigidity coefficient $k$ acquires IR stable point
and the rigidity term is relevant in the IR \cite{Wie89} (see also
\cite{Kav93}).  For $\Theta = 2\pi$ this point corresponds to $k = \frac12$.
So for $\Theta = 2\pi$ we get the fermionic string action considered in
\cite{Wie89,KaKoSe86} for four dimensions and vector representation:

\beqn
S(\Sigma_C) = \eta\int_{\Sigma_C} \sqrt{g} \dd^2 \sigma + \frac12
\int_{\Sigma_C}\sqrt{g}\biggl(\Delta(g) \tilde x_{\mu}\biggr)^2 \dd^2 \sigma
+ \frac12 \int_{\Sigma_C}(e^a_{\mu}\diff_a e^b_{\mu})^2\dd^2 \sigma
+ \nonumber \\ + \frac{i}{16\pi}\int_{\Sigma_C} tr (h^{-1}\dd h)^2 \dd^2
\sigma + \frac{i}{24\pi}\int_B tr (h^{-1}\dd h)^3 \dd^3 s - \ln{J(\tilde x)}
+ O(\frac{1}{m^2}). \label{last}
\eeqn
Due to the existence of IR stable point there is no crumpling of the string
world--sheets in the theory \cite{Dav90}.  Probably there is also no tachyon
in the theory, if crumpling and existence of tachyon are related. Moreover in
the functional integral for the theory \eq{last} there is an integration over
$\cD\tilde x_{\mu}(\sigma)$ which should be defined by the introduction of
the intrinsic metric \cite{Pol87}.  Due to the existence of the IR stable
point of the rigidity coefficient the obtained string theory does not
coincides with the standard Liouville or super Liouville theories
\cite{Pol86}.

   The considered effect of appearance of the spin of the strings from
bosonic theory is general for any string with dyon on it's boundary. This
phenomenon is the mechanism of fermi--bose transmutation for strings in four
dimensions.

   Author is grateful to D.~Dikeman, A.~Gorski, K.~Lee, A.~Losev, A.~Morozov,
M.~Olshanetsky, P.~Orland, V.~Rubakov and especially to M.~Polikarpov for
stimulating discussions and for the help.  This work was supported by the
JSPS Program on Japan -- FSU scientists collaboration, by the Grant
INTAS-94-0840 and by the Grant No.  96-02-17230-a, financed by the RFFS.

\end{document}